\documentclass[twocolumn,showpacs,preprintnumbers,prl,aps,superscriptaddress]{revtex4-2}
\usepackage{graphicx}
\pdfoutput=1
\usepackage{amsmath}
\usepackage{amssymb}
\usepackage[retain-unity-mantissa=false]{siunitx}
\usepackage{xcolor}
\usepackage[caption=false]{subfig}
\usepackage{placeins}
\usepackage[colorlinks=true, linkcolor=blue, citecolor=blue, filecolor=blue, urlcolor=blue]{hyperref}
\usepackage{blindtext}
\usepackage{comment}
\usepackage{verbatim}
\usepackage{ulem}
\usepackage[utf8]{inputenc}
\usepackage[T1]{fontenc} %Make Umlaute great again! (Used for the output
\usepackage{lmodern} %font package
\usepackage{physics}
\usepackage{booktabs} % nicer tables
\usepackage{ragged2e} % ragged right environment and nicer settings
\usepackage{array} % better tables
\usepackage{float}
\usepackage{tikz}
\usetikzlibrary{decorations.pathmorphing}
\usetikzlibrary{decorations.text}
\usetikzlibrary{calc}
\usepackage{svg}
\usepackage{natbib}

\newcommand{\fs}{\femto\second}
\newcommand{\nm}{\nano\meter}

\begin{document}
 
 \title{Competing signatures of intersite and interlayer spin transfer in the ultrafast magnetization dynamics}
 \author{Simon~\surname{Häuser}}
 \email{shaeuser@rptu.de}
 \affiliation{Department of Physics and Research Center OPTIMAS, Rheinland-Pfälzische Technische Universität Kaiserslautern-Landau, 67663 Kaiserslautern, Germany}
 \author{Sebastian~T.\surname{Weber}}
 \affiliation{Department of Physics and Research Center OPTIMAS, Rheinland-Pfälzische Technische Universität Kaiserslautern-Landau, 67663 Kaiserslautern, Germany}
 \author{Christopher~\surname{Seibel}}
 \affiliation{Department of Physics and Research Center OPTIMAS, Rheinland-Pfälzische Technische Universität Kaiserslautern-Landau, 67663 Kaiserslautern, Germany}
 \author{Marius~\surname{Weber}}
 \affiliation{Department of Physics and Research Center OPTIMAS, Rheinland-Pfälzische Technische Universität Kaiserslautern-Landau, 67663 Kaiserslautern, Germany}
 \author{Laura~\surname{Scheuer}}
 \affiliation{Department of Physics and Research Center OPTIMAS, Rheinland-Pfälzische Technische Universität Kaiserslautern-Landau, 67663 Kaiserslautern, Germany}
 \author{Martin~\surname{Anstett}}
 \affiliation{Department of Physics and Research Center OPTIMAS, Rheinland-Pfälzische Technische Universität Kaiserslautern-Landau, 67663 Kaiserslautern, Germany}
 \author{Gregor~\surname{Zinke}}
 \affiliation{Department of Physics and Research Center OPTIMAS, Rheinland-Pfälzische Technische Universität Kaiserslautern-Landau, 67663 Kaiserslautern, Germany}
 \author{Philipp~\surname{Pirro}}
 \affiliation{Department of Physics and Research Center OPTIMAS, Rheinland-Pfälzische Technische Universität Kaiserslautern-Landau, 67663 Kaiserslautern, Germany}
 \author{Burkard~\surname{Hillebrands}}
 \affiliation{Department of Physics and Research Center OPTIMAS, Rheinland-Pfälzische Technische Universität Kaiserslautern-Landau, 67663 Kaiserslautern, Germany}
 \author{Hans Christian~\surname{Schneider}}
 \affiliation{Department of Physics and Research Center OPTIMAS, Rheinland-Pfälzische Technische Universität Kaiserslautern-Landau, 67663 Kaiserslautern, Germany}
 \author{Bärbel~\surname{Rethfeld}}
 \affiliation{Department of Physics and Research Center OPTIMAS, Rheinland-Pfälzische Technische Universität Kaiserslautern-Landau, 67663 Kaiserslautern, Germany}
 \author{Benjamin~\surname{Stadtmüller}}
 \email{b.stadtmueller@rptu.de}
 \affiliation{Department of Physics and Research Center OPTIMAS, Rheinland-Pfälzische Technische Universität Kaiserslautern-Landau, 67663 Kaiserslautern, Germany}
 \affiliation{Institute of Physics, Johannes Gutenberg University Mainz, 55128 Mainz, Germany}
 \author{Martin~\surname{Aeschlimann}}
 \affiliation{Department of Physics and Research Center OPTIMAS, Rheinland-Pfälzische Technische Universität Kaiserslautern-Landau, 67663 Kaiserslautern, Germany}
 \date{\today}
 
    \begin{abstract}
 	Optically driven intersite and interlayer spin transfer are individually known as the fastest processes for manipulating the spin order of magnetic materials on the sub $100\,$fs time scale. However, their competing influence on the ultrafast magnetization dynamics remains unexplored. In our work, we show that optically induced intersite spin transfer (also known as OISTR) dominates the ultrafast magnetization dynamics of ferromagnetic alloys such as Permalloy (Ni\textsubscript{80}Fe\textsubscript{20}) only in the absence of interlayer spin transfer into a substrate. Once interlayer spin transfer is possible, the influence of OISTR is significantly reduced and interlayer spin transfer dominates the ultrafast magnetization dynamics. This provides a new approach to control the magnetization dynamics of alloys on extremely short time scales by fine-tuning the interlayer spin transfer.
 \end{abstract}
 
 \maketitle

%%%%%%%%%%%%%%%%%%%%%%%%%%  introduction

Increasing the operating speed of modern spintronic devices requires new strategies to manipulate, transport, and store digital information encoded in the spin angular momentum of electrons on shorter time scales. One promising way to achieve this goal is to use ultrashort femtosecond (fs) light pulses as external stimuli to modify the material properties of spintronic relevant materials, such as ferromagnets and antiferromagnets. The feasibility of this approach for the realization of ultrafast spintronics has been demonstrated by pioneering studies in the last two decades, which revealed the (sub-) picosecond loss of magnetic order in ferro- and antiferromagnets after excitation with ultrashort optical and THz pulses \cite{Beaurepaire1996, Aeschlimann1997,Scholl1997,Hohlfeld1997, Guedde1999, Koopmans2000, Regensburger2000,Qiu2020,Schmidt2010,Bossini2016,Kampfrath2010} and even reported the all optical magnetization reversal by fs light pulses \cite{Stanciu2007,Radu2011,Lambert2014}. In most cases, however, the time scale of the material response and the corresponding change in the spin order of the material is not related to the duration of the optical excitation itself (i.e., the length of the fs-light pulse). Instead, the magnetization dynamics evolves on a significantly longer intrinsic time scale that is governed by secondary angular momentum dissipation processes such as electron-electron scattering \cite{Krauss2009, Mueller2013}, Elliot-Yafet electron-phonon spin flip scattering \cite{Koopmans2009,Koopmans2005,Carva2011,Mueller2013}, generation of ultrafast non-coherent magnons \cite{Schmidt2010, Eich2017}. These processes limit the optical material response in metallic 3d ferromagnets, such as nickel, to  $\approx 100\,$fs \cite{Beaurepaire1996,Koopmans2005,Koopmans2009}.

Faster material responses have only been reported for magnetic materials where the ultrafast magnetization dynamics are dominated, or at least significantly influenced, by optically induced spin transport and transfer processes. One of these processes is the superdiffusive spin transport \cite{Battiato2010,Hofherr2017,Shokeen2017,Eschenlohr2013} in magnetic bilayer and multilayer structures. In this case, the different velocities of the optically excited minority and majority carriers in the ferromagnet lead to an effective ultrafast transport of spin angular momentum out of the magnetic material into an adjacent layer and thus to an ultrafast demagnetization that can be faster than $50\,$fs \cite{Hofherr2017}. The second spin transfer process relevant on these ultrashort time scales is the so-called optical intersite spin transfer (OISTR) \cite{Dewhurst2018}. It refers to an optically induced spin transfer between different magnetic subsystems of a magnetic alloy or multilayer system, which is purely mediated by the optical transitions between different electronic states of the materials \cite{hofherr2020,Willems2020,Tengdin2020, Siegrist2019,Chen2019,Steil2020,Golias2021}. As a result, the influence of OISTR on the magnetization dynamics of materials is determined solely by the pulse length of the optical excitation.

In reality, however, OISTR and superdiffusive spin transport influence the demagnetization dynamics on nearly identical time scales. This is mainly due to the fact that the fs light pulses used to manipulate magnetic materials are typically generated with pulse durations in the range between $20\,$fs and $50\,$fs. This could potentially lead to a competing influence of OISTR and superdiffusive spin transport on the ultrafast magnetization dynamics of magnetic multilayer structures. Understanding and exploiting this competition thus offers an intriguing pathway to optically engineer the ultrafast magnetization dynamics on the sub $100\,$fs time scale. 

In this work we investigate the competing influence of OISTR and interlayer spin transfer via superdiffusive spin transport on the ultrafast magnetization dynamics of the ferromagnetic alloy Permalloy (Py, Ni\textsubscript{80}Fe\textsubscript{20}). Tuning the substrate material of the Py films from the insulator MgO to gold films of different thicknesses (\SI{10}{\nano\meter} and \SI{100}{\nano\meter}) allows us to gradually increase the role of interlayer spin transfer for the ultrafast demagnetization dynamics. Our joint experimental and theoretical work provides substantial evidence that the OISTR signature and thus the relevance of OISTR for the demagnetization dynamics of Py decreases with increasing importance of interlayer spin transfer from Py into the adjacent metallic layer. Our results underscore the competing roles of intralayer and interlayer spin transfer for the ultrafast magnetization dynamics of magnetic multilayer structures on the ultrafast, sub $100\,$fs time scale.

%%%%%%%%%%%%%%%%%%%%%%%%%%  experiment

In our study, we investigate three Permalloy (Ni\textsubscript{80}Fe\textsubscript{20}) thin films (\SI{10}{\nano\meter}) deposited on different substrates: (i) directly on the insulating substrate MgO, (ii) on a \SI{10}{\nano\meter} Au film on MgO, and (iii) on a \SI{100}{\nano\meter} Au film. All samples were protected from oxidation by a \SI{2}{\nano\meter} Al\textsubscript{2}O\textsubscript{3} capping layer.\\
The time- and element-resolved magnetization dynamics of these samples were investigated by time-resolved Kerr spectroscopy with fs- extreme UV (XUV) radiation in transverse geometry (T-MOKE). Using neon for the generation of the fs XUV radiation by high harmonic generation (HHG) (similar to Ref.~\cite{OVorakiat2012}), we can cover a spectral range of $40-72\,$eV that coincides with the characteristic M\textsubscript{2,3} absorption edges of Fe and Ni at \textasciitilde \SI{52.7}{\electronvolt} and \textasciitilde \SI{66.2}{\electronvolt} \cite{xraybooklet}, respectively. 
\begin{figure}[t]
	\centering
	\includegraphics[scale=1]{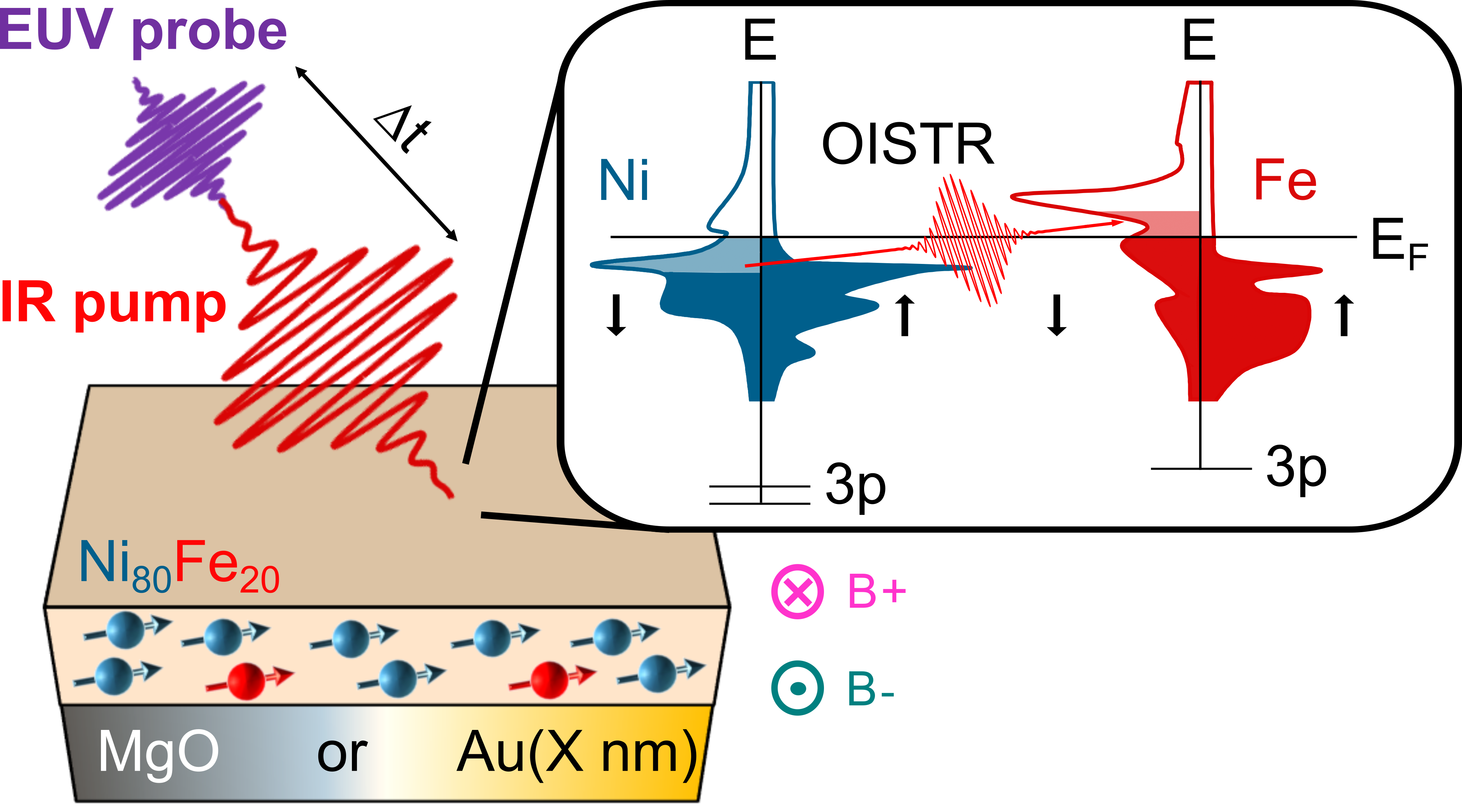}
	\caption{(Left) Sketch of the time-resolved Kerr spectroscopy experiment. An optical pump pulse ($1.55\,$eV, $~30\,$fs) excites the sample while the element specific magnetization dynamics are monitored by changes in the magnetic asymmetry at the Fe M\textsubscript{2,3} and Ni M\textsubscript{3} absorption edges. (Right) Schematic representation of the density of states of Fe and Ni in Py and the population changes due to the OISTR. Optical excitation by the IR pump leads to an effective spin transfer from the occupied Ni minority channel to the Fe minority channel.}
	\label{figure:measurement}
\end{figure}

The magnetic contrast is obtained from the absorption spectra by recording the reflectivity of the whole fs-XUV spectrum of the sample for two alternating directions of the magnetic $B$-field $I_{+}$ and $I_{-}$ as shown in \autoref{figure:measurement}. Then, we calculate the magnetic asymmetry $A$ by \cite{Valencia2006, OVorakiat2012} 
\begin{equation}
	\label{eqn:asymmetry}
	A = \frac{I_{+} - I_{-}}{I_{+} + I_{-}} \propto \text{Spin Polarization}.
\end{equation}
The energy resolved asymmetry $A$ is proportional to the spin polarization (SP) \cite{Erskine1975,Hoechst1997} and gives clear signatures of OISTR as shown by Hofherr et al. \cite{hofherr2020}.

For the optical excitation of our sample, we used a laser fluence of 28.1\,mJ/cm\textsuperscript{2}, which resulted in a different loss of magnetization for each sample. Therefore, we applied a dedicated normalization procedure to all magnetization traces using the data for Py/Au(10\,nm) as a reference.

%%%%%%%%%%%%%%%%%%%%%%%%%%  results_and_discussion

\begin{figure}[h]
	\centering
	\includegraphics[scale=1]{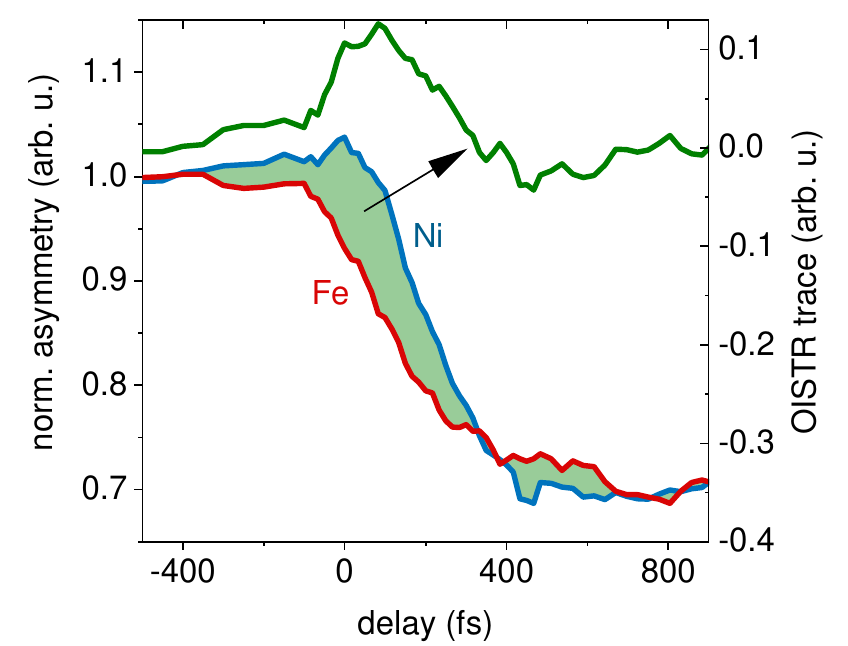}
	\caption{Temporal evolution of the spin polarization of the corresponding Ni states (blue line) below and Fe states (red line) above $E_{F}$ for Py/MgO. The difference between the spin polarization of the Ni and Fe states is shown as a green curve.}
	\label{fig:OISTR_delta}
\end{figure}

We start our discussion with the magnetization dynamics of Py/MgO where the initial magnetization dynamics are dominated by OISTR and interlayer spin transfer into the substrate is absent. As reported for another FeNi alloy\cite{hofherr2020}, the optical excitation by the IR pump pulse leads mainly to a transfer from the minority electrons of Ni into the unoccupied minority states of Fe close to the Fermi energy $E_{F}$, as shown in the inset of \autoref{figure:measurement}.
This intralayer spin transfer can be recorded and visualized in the T-MOKE experiment by extracting the time-dependent spin polarization for selected spectral regions in close vicinity to the M\textsubscript{2,3} absorption edges of Ni (\textasciitilde$66.2\,$eV) and Fe (\textasciitilde$52.7\,$eV). In particular, we follow our previous work \cite{hofherr2020} and evaluate the spin polarization for an energy range corresponding to the occupied Ni states below the Fermi energy and the unoccupied Fe states located slightly above $E_{F}$. The corresponding traces are shown as blue (Ni states) and red (Fe states) curves in \autoref{fig:OISTR_delta}.

The SP of the Ni states increases instantaneously upon laser excitation, coinciding with a decrease in the SP of the Fe states. This opposite behavior has been reported as the spectroscopic OISTR signature and characterizes the initial influence of OISTR on the ultrafast magnetization dynamics of Py. It can be further quantified by the so-called OISTR trace (OT), i.e. the difference between the SP of the Ni and Fe states, shown as a green curve in \autoref{fig:OISTR_delta}. The OT reaches its maximum after about \SI{200}{\fs} and disappears again within the next \SI{250}{\fs}. The rise of the OT can be directly related to the OISTR and the corresponding initial changes in the magnetization of the Fe and Ni sublattices. The decay of the OT is attributed to exchange scattering \cite{Mathias2012,Muenzenberg2010} and spin flip scattering processes occurring locally within the Py film.

\begin{figure}[H]
	\centering
	\includegraphics[scale=1]{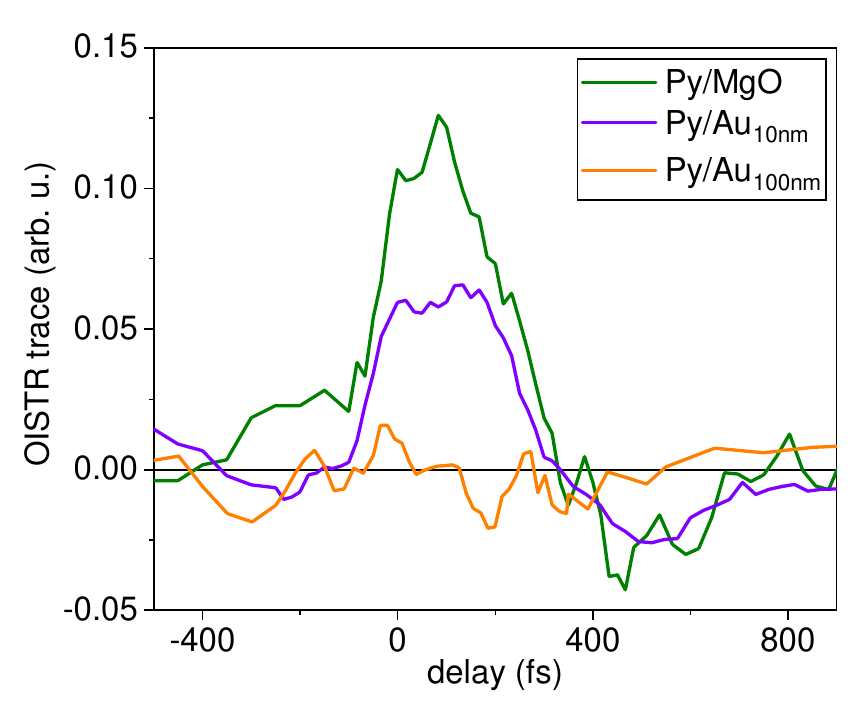}
	\caption{Temporal evolution of the OT for Py on three different substrates: insulating MgO (green curve), metallic \SI{10}{\nm} Au film, and \SI{10}{\nm} Au film.}
	\label{fig:allOISTR_deltas}
\end{figure}

To explore the competing influence of OISTR and interlayer spin transport on the ultrafast magnetization dynamics of Py, we now turn to the OT for similar Py films on metallic Au films of 10\,nm and 100\,nm thickness. We recorded similar time-resolved T-MOKE data sets for both samples (see Supplementary Material) and determined the OT using the identical procedure as described above for the Py film on MgO. The resulting OTs are summarized for all three sample systems in \autoref{fig:allOISTR_deltas}. We find a clear decrease in the magnitude of the OT for Py/Au(\SI{10}{\nm}) compared to Py/MgO. More importantly, the OT disappears completely for Py on the \SI{100}{\nm} Au film.

Our observation clearly demonstrates that the initial changes in spin polarization and the magnetization dynamics due to OISTR are greatly reduced or even suppressed by replacing the insulating substrate with a metallic thin film. Therefore, we propose that these changes in OT are due to interlayer spin transfer and transport, rather than a modification of the spin flip scattering rate in Py due to the hybridization of Py and Au at the interface. In particular, no local interfacial effect could account for the observed thickness dependent changes in the OT.

\begin{figure}[h]
	\centering
	\includegraphics[scale=1.0]{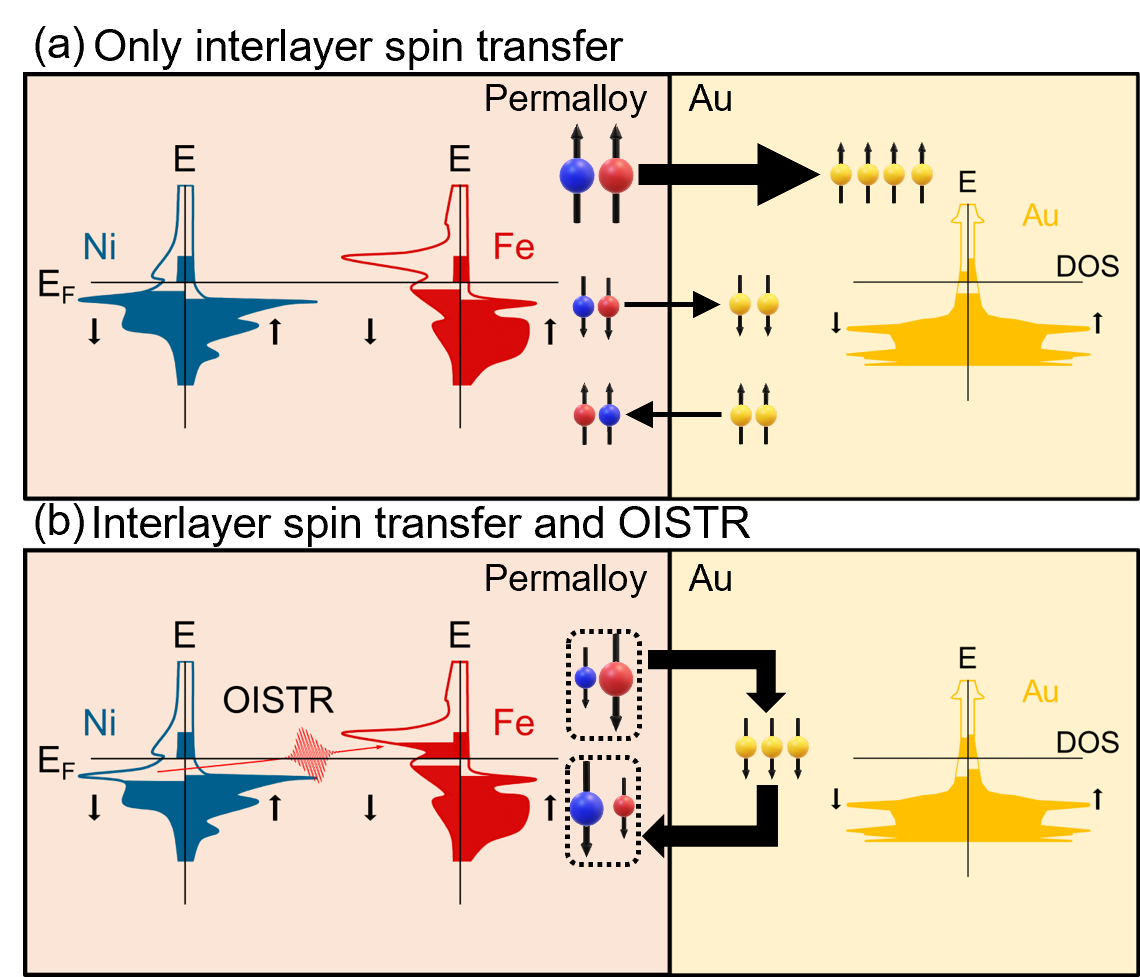}
	\caption{Schematic illustration of the different optically induced interlayer spin transfer pathways of majority and minority carriers across the Py/Au interface. The hypothetical flow of spin-polarized carriers in the absence of OISTR is shown in panel (a), the additional interlayer spin transfer pathways due to OISTR are highlighted in panel (b).}
	\label{fig:OISTRtracedecaycurrent}
\end{figure} 

\autoref{fig:OISTRtracedecaycurrent} outlines a possible scenario explaining the reduction of OT in the Py film by interlayer spin transfer:
In the absence of OISTR, interlayer spin transfer leads to a flow of spin-polarized charges from the ferromagnetic material into the non-magnetic substrate and a corresponding backflow of unpolarized charges to maintain charge neutrality. This is shown in \autoref{fig:OISTRtracedecaycurrent}(a). The magnitude of the interlayer spin transfer for both majority and minority carriers depends (apart from the interfacial transmission efficiency), on the number of carriers in the ferromagnetic material, i.e. in Py, and the number of available states in the substrate material, i.e. in Au. This results in an overall majority carrier dominated spin transfer from Py to Au. 
In the case of OISTR, the optical excitation additionally redistributes minority electrons from Ni states below $E_{F}$ to Fe states above $E_{F}$, as shown on the right side of \autoref{figure:measurement}. Thus, OISTR leads to a significant change in the number of carriers and final states available for interlayer spin transfer. This population change leads to additional interlayer spin transfer channels as shown in \autoref{fig:OISTRtracedecaycurrent}(b). The optically excited Fe minority electrons can now travel through the interface, reducing the excited state spin population initially created by the OISTR. On the other hand, the backflow of electrons from Au to Py can repopulate the minority states in Ni that were initially depopulated by the OISTR. In this way, both additional transport channels would substantially counteract the changes in the spin-dependent population created by the OISTR and thus reduce the OT observed in our experiment. 
However, our model does not yet account for the increasing reduction in OT magnitude with increasing film thickness.

%%%%%%%%%%%%%%%%%%%%%%%%%%   theory

\begin{figure}
	\centering
	\includegraphics{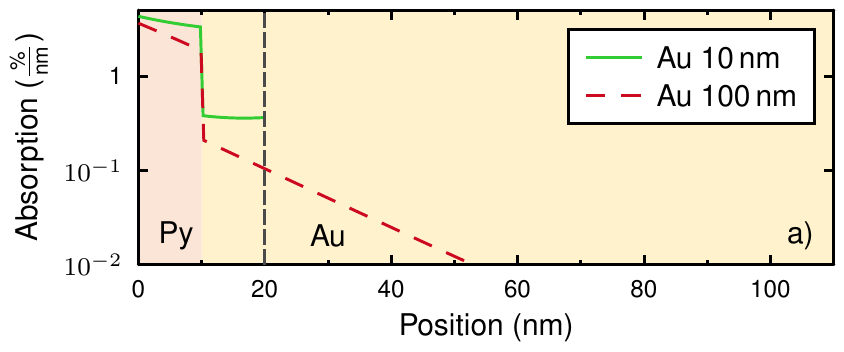}
	\includegraphics{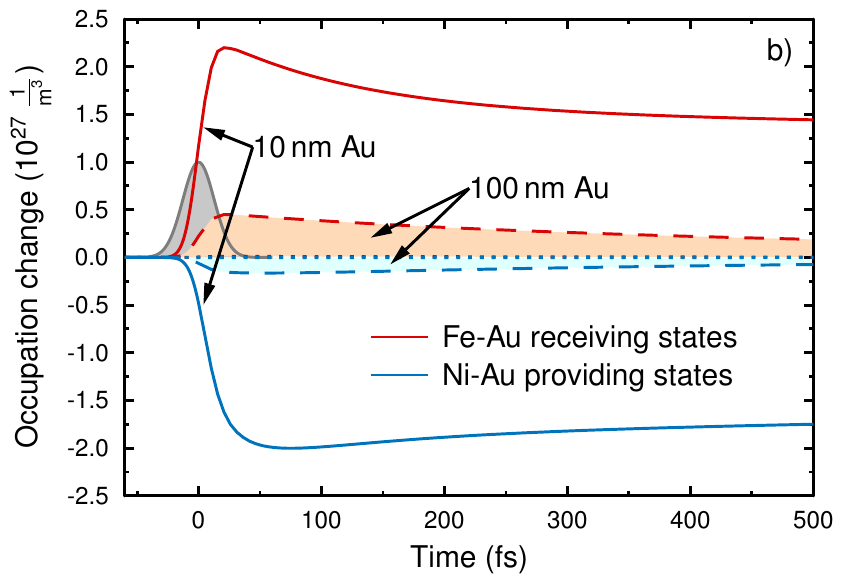}
	\caption{a) Structure of the samples and absorption profiles. b) Change of the occupation due to excitation for states slightly above Fermi energy (red lines) and slightly below Fermi energy (blue lines) in Au, respectively. The solid lines show the occupation difference for the \SI{10}{\nano\meter} Au layer. 
		The shaded areas mark the occupation difference for the case of the \SI{100}{\nano\meter} Au layer for different assumptions on laser-energy transport (see text). 
		The temporal shape of the laser pulse is shown in gray.}
	\label{fig:occupation_change}
\end{figure}

To adress this final question theoretically, we model the ultrafast population of states around the Fermi energy in Au with the help of a kinetic approach considering the optical excitation as well as the electron-electron and electron-phonon interactions, each with a full Boltzmann collision integral in dependence of time and energy \cite{Mueller2013PRB}. This allows us to determine the non-equilibrium population and depopulation of states on the same time scales as the measured OISTR signal in the Py layer. 
The excitation strength is determined from the experimental fluence and absorption calculations for the experimental multilayer including capping and substrate with refractive indices from Ref.~\cite{Tikuisis2017}.  
The resulting absorption profiles for a \SI{10}{\nm} Au-film and for a \SI{100}{\nm} Au-film are shown in \autoref{fig:occupation_change}~a). The dashed line marks the backside of the \SI{10}{\nm} Au-film, which is responsible for the back reflections, leading to a much larger total absorption in the thinner Au film as compared to the thicker film, note the logarithmic scale in \autoref{fig:occupation_change}~a). 
Furthermore, we assume a homogeneous distribution of the absorbed energy for the thin film of 10\,nm thickness.  
This is already roughly fulfilled by the 
spatial absorption profile depicted in \autoref{fig:occupation_change}\,a), 
but also because
the range of ultrafast homogeneous energy distribution by ballistic electrons has been estimated to be about \SI{100}{\nm} for Au \cite{Hohlfeld1997}, which is much larger than the thin Au substrate.
In contrast, for the 100\,nm Au-film, we determine two different limits of absorption strength based on the profile shown in  \autoref{fig:occupation_change}\,a), as will be described later. 
With the given amount of absorbed energy we model the laser-excitation of electrons in Au. 
Typical resulting non-equilibrium electron distributions in noble metals can be found, e.g., in Refs.~\cite{Mueller2013PRB,Weber2017,Weber2019}. 

We evaluate the change of occupation in the states at $E= E_F - \SI{1.8}{\electronvolt}$ and $E= E_F + \SI{1.35}{\electronvolt}$
according to experimentally measured energy regions. The width of the evaluated interval, $\Delta E = \SI{0.7}{\electronvolt} $, is chosen corresponding to the width of the HHG peaks of the experiment.
As indicated in \autoref{fig:OISTRtracedecaycurrent}, the energy window above the Fermi energy $E_F$ corresponds to Fe states providing minority carriers for the transport into the Au substrate, while the energy below the Fermi energy is relevant for receiving minority carriers from Au in Ni.

The time evolution of the occupation is given as
\begin{equation}
	n(t) = \int\displaylimits_{E-\Delta E/2}^{E+\Delta E/2} f(\epsilon, t) D(\epsilon) \mathrm{d}\epsilon\,,
\end{equation}
where $f$ denotes the electronic distribution function and $D$ its density of states (DOS). 

The change of the occupation $n(t)$ as compared to its initial value at room temperature is shown in 
\autoref{fig:occupation_change}\,b). 
In the case of the \SI{10}{\nano\meter} Au layer, the excitation leads to a fast increase of the occupation above $E_{F}$  and a decrease below $E_{F}$ within the time scale of the laser pulse.
After the laser pulse, relaxation processes lead to a decrease of both signals. 

This shows that the laser excitation is blocking the states relevant for non-local transport across the interface effectively on the time scale of the pump pulse duration and the subsequent several \SI{100}{\femto\second}, see solid lines in \autoref{fig:occupation_change}\,b). 
Thus, the transport scenario depicted in \autoref{fig:OISTRtracedecaycurrent} is partially blocked and while the measured OISTR signal in Py is reduced as compared to an insulating substrate, but it does not vanish completely.

In the case of the thicker \SI{100}{\nano\meter} Au layer, the spin transport from Py to Au is not blocked, leading to a complete extinction of the resolvable OISTR signal. 
We show this for two limiting cases of Au excitation. In a first calculation, we assume that the fraction of pump energy absorbed in Au, see \autoref{fig:occupation_change}\,a), is homogeneously distributed by ballistic transport over the \SI{100}{\nano\meter} Au layer.
The occupation change in the relevant Au states is not resolved in \autoref{fig:occupation_change}~b), see dotted lines. Thus, in this case, Au can act as a spin acceptor or source, respectively, and thus eliminate the OT. 
To evaluate the possibility of state-blocking in the \SI{100}{\nano\meter} Au substrate as we found for the thin \SI{10}{\nano\meter} Au film, we 
consider no energy transport within the Au layer and average just over the first \SI{10}{\nano\meter} closest to the Py interface. 
The excitation of this virtual layer is lower than in the case of the \SI{10}{\nano\meter} Au film, because no back reflection effects are active in the bulk substrate.
The resulting occupation changes within the energy intervals relevant for spin transport 
are shown by the dashed lines in \autoref{fig:occupation_change}~b), which represent the maximum occupation change caused by the given excitation parameters. 
While these occupation changes are qualitatively similar to the corresponding changes we determined for the thin film (solid lines), their magnitude is much weaker. 
Thus, the spin current from Py into Au is not blocked by the excited non-equilibrium electron distribution and efficiently extincts the OISTR signal. 

Our theoretical model thus explains the existence of an OISTR signal in Py on a metallic substrate by a blocked spin transport across the interface due to a transient non-equilibrium electron distribution in the metallic substrate. 
This state blocking is more effective at higher absorption strength in the metal. 
For a thin, \SI{10}{\nano\meter} Au film, we find a higher absorption due to multireflection in the thin film. 
Moreover, the absorbed energy cannot be dissipated to the depth by ballistic energy transport as it is known for gold films \cite{Hohlfeld1997}, but remains confined in this thin layer close to the experimentally studied Py film.

%%%%%%%%%%%%%%%%%%%%%%%%%%   conclusion

In conclusion, we uncover the competing influence of OISTR and interlayer spin transfer on the initial ultrafast magnetization dynamics of the ferromagnetic alloy Permalloy. We show that OISTR dominates the ultrafast magnetization dynamics of the alloy only in the absence of interlayer spin transfer into a metallic substrate. However, once interlayer spin transfer is possible, the optically redistributed spins due to OISTR are, at least partially, transferred from the alloy into the metallic substrate and thus no longer dominate the initial ultrafast demagnetization dynamics. Thus, our study demonstrates a clear way to tune the competing influence of OISTR and interlayer spin transfer on the magnetization dynamics by controlling the efficiency of interlayer spin transfer across interfaces.

%%%%%%%%%%%%%%%%%%%%%%%%%%   acknowledgement 

\begin{acknowledgements}	

The experimental work was funded by the Deutsche Forschungsgemeinschaft (DFG, German Research Foundation) - TRR 173 - 268565370 Spin + X: spin in its collective environment (Project A08 and B11). B.S. further acknowledges funding by the Dynamics and Topology Research Center (TopDyn) funded by the State of Rhineland Palatinate.

\end{acknowledgements}

\end{document}

\newcommand{\gpi}{\textrm{\greektext p}}
	\newcommand{\gmu}{\textrm{\greektext m}}
	\newcommand{\beginsupplement}{%
		\setcounter{table}{0}
		\renewcommand{\thetable}{S\arabic{table}}%
		\setcounter{figure}{0}
		\renewcommand{\thefigure}{S\arabic{figure}}%
		\renewcommand{\theequation}{S\arabic{equation}}
	    }

	\title{Competing signatures of intersite and interlayer spin transfer in the ultrafast magnetization dynamics - Supplemental Material}
	\author{Simon~\surname{Häuser}}
	\email{shaeuser@rptu.de}
	\affiliation{Department of Physics and Research Center OPTIMAS, Rheinland-Pfälzische Technische Universität Kaiserslautern-Landau, 67663 Kaiserslautern, Germany}
	\author{Sebastian T.~\surname{Weber}}
	\affiliation{Department of Physics and Research Center OPTIMAS, Rheinland-Pfälzische Technische Universität Kaiserslautern-Landau, 67663 Kaiserslautern, Germany}
	\author{Christopher~\surname{Seibel}}
	\affiliation{Department of Physics and Research Center OPTIMAS, Rheinland-Pfälzische Technische Universität Kaiserslautern-Landau, 67663 Kaiserslautern, Germany}
	\author{Marius~\surname{Weber}}
	\affiliation{Department of Physics and Research Center OPTIMAS, Rheinland-Pfälzische Technische Universität Kaiserslautern-Landau, 67663 Kaiserslautern, Germany}
	\author{Laura~\surname{Scheuer}}
	\affiliation{Department of Physics and Research Center OPTIMAS, Rheinland-Pfälzische Technische Universität Kaiserslautern-Landau, 67663 Kaiserslautern, Germany}
	\author{Martin~\surname{Anstett}}
	\affiliation{Department of Physics and Research Center OPTIMAS, Rheinland-Pfälzische Technische Universität Kaiserslautern-Landau, 67663 Kaiserslautern, Germany}
	\author{Gregor~\surname{Zinke}}
	\affiliation{Department of Physics and Research Center OPTIMAS, Rheinland-Pfälzische Technische Universität Kaiserslautern-Landau, 67663 Kaiserslautern, Germany}
	\author{Philipp~\surname{Pirro}}
	\affiliation{Department of Physics and Research Center OPTIMAS, Rheinland-Pfälzische Technische Universität Kaiserslautern-Landau, 67663 Kaiserslautern, Germany}
	\author{Burkard~\surname{Hillebrands}}
	\affiliation{Department of Physics and Research Center OPTIMAS, Rheinland-Pfälzische Technische Universität Kaiserslautern-Landau, 67663 Kaiserslautern, Germany}
	\author{Hans Christian~\surname{Schneider}}
	\affiliation{Department of Physics and Research Center OPTIMAS, Rheinland-Pfälzische Technische Universität Kaiserslautern-Landau, 67663 Kaiserslautern, Germany}
	\author{Bärbel~\surname{Rethfeld}}
	\affiliation{Department of Physics and Research Center OPTIMAS, Rheinland-Pfälzische Technische Universität Kaiserslautern-Landau, 67663 Kaiserslautern, Germany}
	\author{Benjamin~\surname{Stadtmüller}}
	\email{b.stadtmueller@rptu.de}
	\affiliation{Department of Physics and Research Center OPTIMAS, Rheinland-Pfälzische Technische Universität Kaiserslautern-Landau, 67663 Kaiserslautern, Germany}
	\affiliation{Institute of Physics, Johannes Gutenberg University Mainz, 55128 Mainz, Germany}
	\author{Martin~\surname{Aeschlimann}}
	\affiliation{Department of Physics and Research Center OPTIMAS, Rheinland-Pfälzische Technische Universität Kaiserslautern-Landau, 67663 Kaiserslautern, Germany}
	
	\date{\today}

	\maketitle
	\beginsupplement

	\section{Sample Preparation}
	
    Py thin films and Py/Au bilayers were grown on \SI{0.5}{\milli\meter}-thick MgO\,(100) by molecular beam epitaxy (MBE) technique in an ultrahigh vacuum (UHV) chamber at a base pressure of \SI{5E-9}{\milli\bar}. The cleaning protocol of the MgO substrates included ex-situ chemical cleaning with acetone and isopropanol and in-situ heating at
    \SI{600}{\celsius} for \SI{1}{\hour}. Py deposition was performed at room temperature with a growth rate in the range of
    $R=0.15$\,\AA/s ($R=0.15$\,nm/s) controlled by a quartz crystal oscillator. In the case of the bilayer structures, a \SI{10}{\nano\meter} or \SI{100}{\nano\meter} Au film was first grown directly on the MgO substrate using similar parameters.

	\section{Experimental details} \label{exp.details}
	
    Our time-resolved experiments were performed with a high harmonic generation transversal magneto optical Kerr effect (HHG-T-MOKE) pump-probe setup \cite{OVorakiat2012, Mathias2012}. An amplified Ti:sapphire laser system (1.57\,eV, 30\,fs, 6\,kHz, 1.7\,mJ per pulse) was used to generate the pump and probe beams. To generate the XUV probe beam, we used the high harmonic generation process with a fiber based setup and neon as the excited noble gas  \cite{OVorakiat2012, Mathias2012}. We determined the temporal pulse length at the sample position to 80\,fs for the 1.57\,eV pump pulse using an autocorrelation method. 
    Both p-polarized pulses were focused on the sample at an angle of incidence of $45$ degrees. The reflected XUV light is separated from the residual fundamental light of 1.57\,eV by two 100\,nm aluminum filters and then detected by a spectrometer to ensure energy and therefore element resolution. The estimated energy resolution is $\approx$\,0.8\,eV.

	\section{Spectral ranges for evaluation} \label{spectral evaluation}
	\vspace{-18pt}
	\begin{figure}[H]
		\centering
		\subfloat{\includegraphics[scale=1]{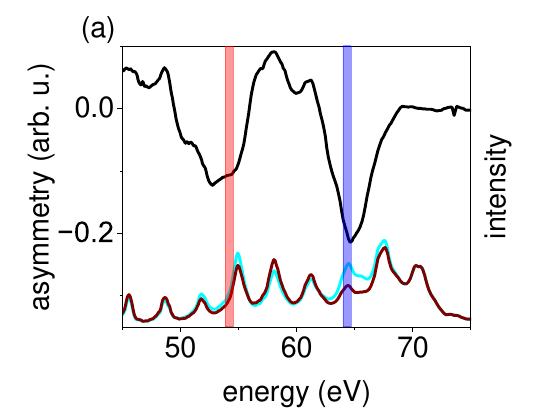}}
		\subfloat{\includegraphics[scale=1]{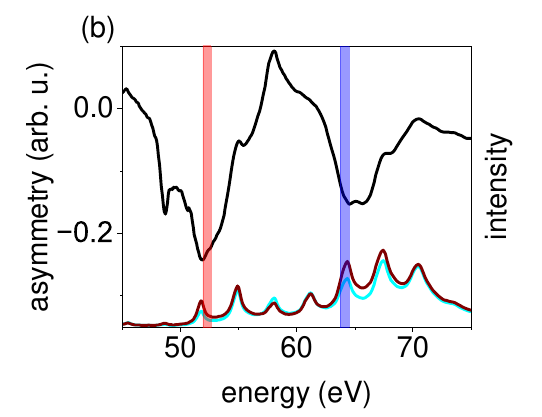}}
		\subfloat{\includegraphics[scale=1]{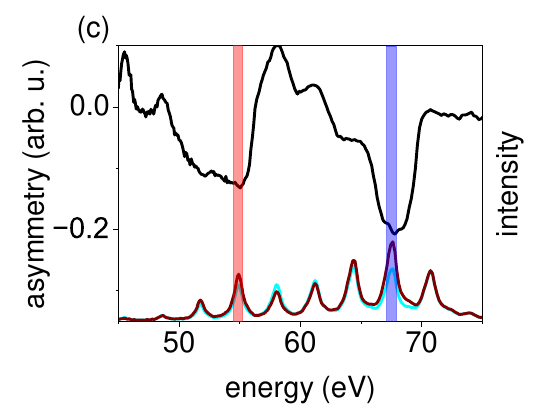}}
		\caption{Detected XUV-spectra (cyan and magenta) and the resulting asymmetries of Py on (a) MgO, (b) \SI{10}{\nano\meter}\,Au and (c) \SI{100}{\nano\meter}\,Au. The red and blue shaded areas correspond to the evaluated spectral regions for Fe and Ni, respectively.}
		\label{fig:}
	\end{figure}

	\section{Scaling of magnetization dynamics}

\begin{figure}[h]
	\centering
	\includegraphics[scale=1]{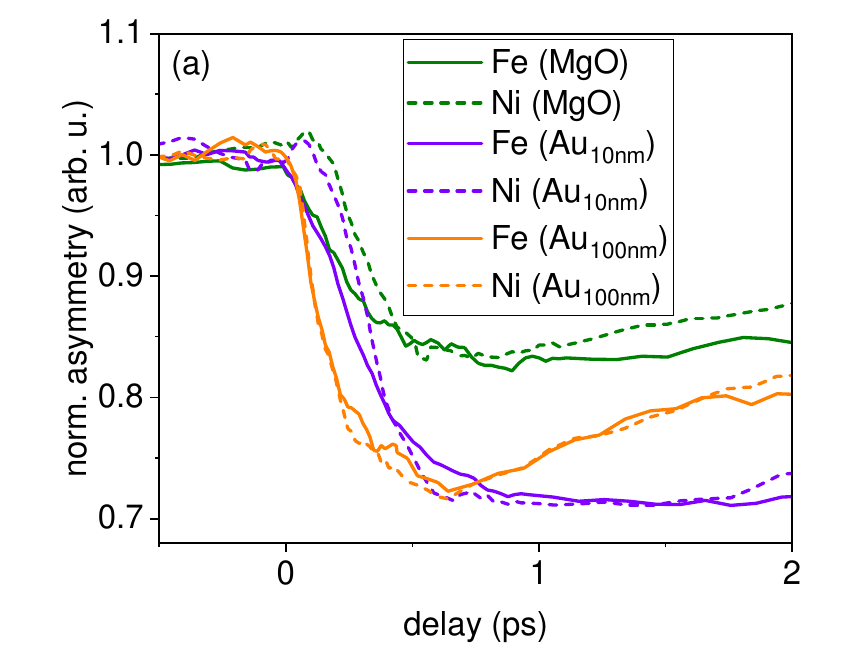}
	\includegraphics[scale=1]{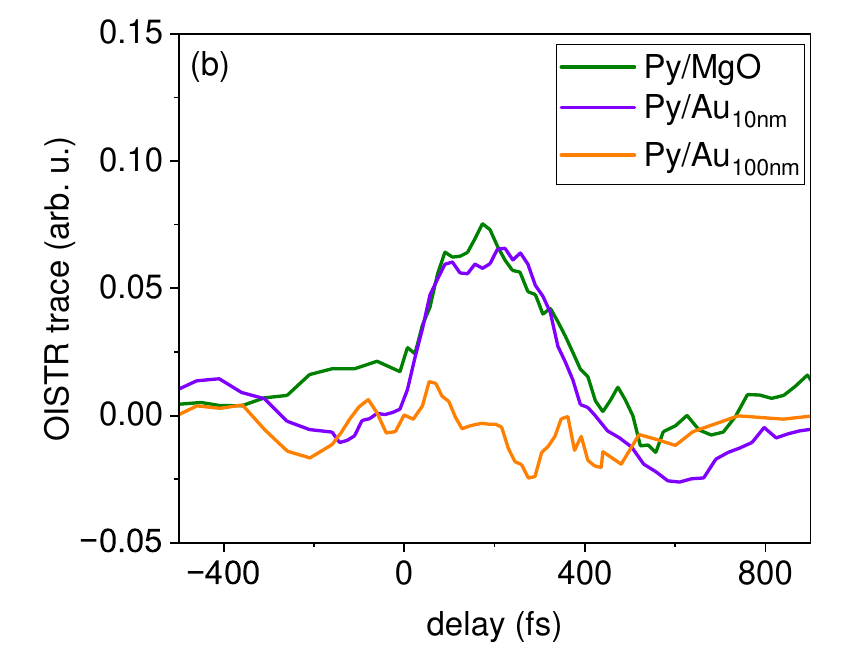}
	\includegraphics[scale=1]{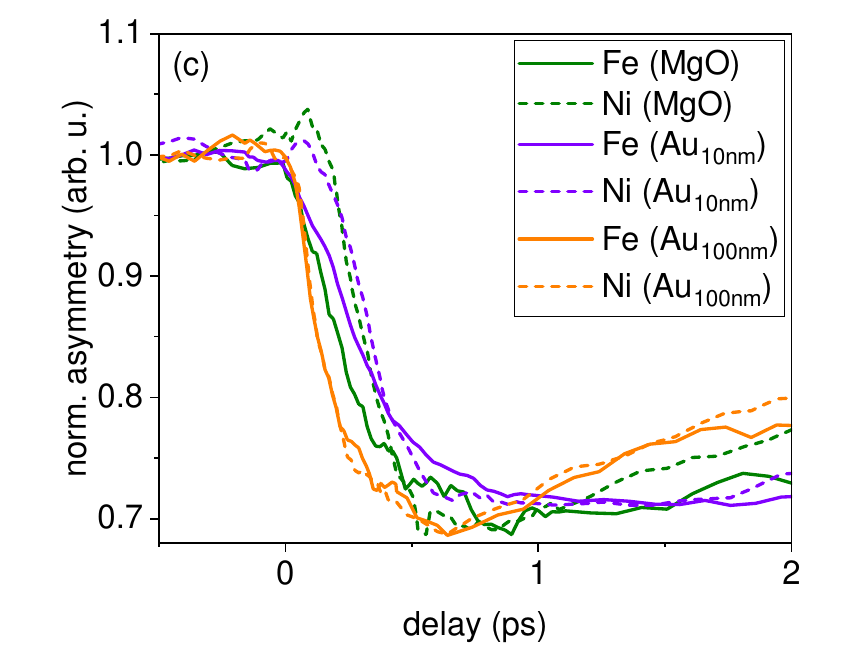}
	\includegraphics[scale=1]{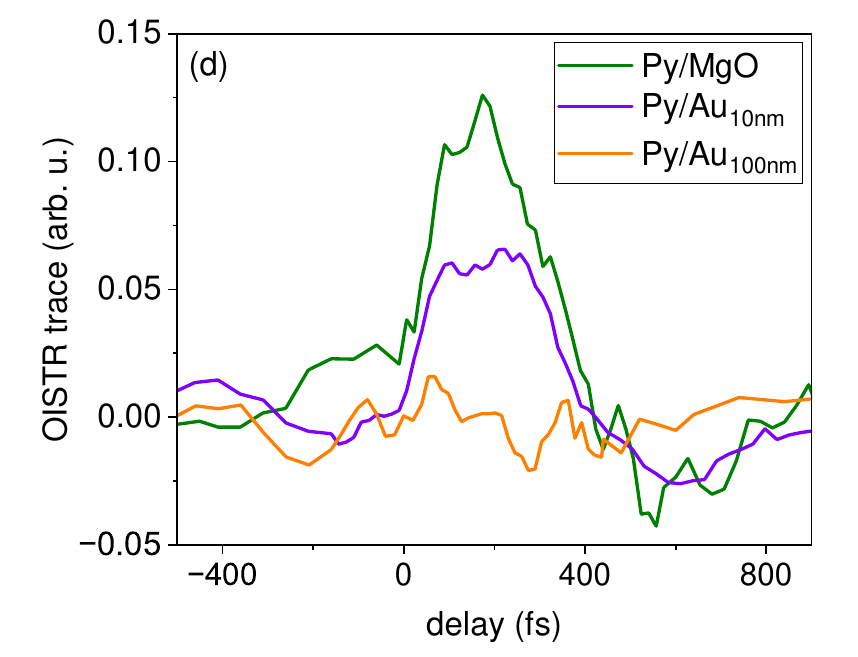}
	\caption{(a) Raw dynamic of the spin polarization of Py on MgO, Au(10\,nm) and Au(100\,nm). (b) Resulting OISTR trace of (a). (c) Dynamic of the spin polarization of Py on MgO, Au(10\,nm) and Au(100\,nm) normalized to $\approx$\,30\% quenching. (d) Resulting OISTR trace of (c).}
	\label{fig:demags_and_OT_scaling}
\end{figure}

As mentioned in the experimental details of the paper, we applied the same fluence for each Py sample with a different substrate, resulting in different quenching. These results are shown in Fig.~\ref{fig:demags_and_OT_scaling}\,(a). Here, the Py/Au(10\,nm) and Py/Au(100\,nm) sample systems lost almost the same amount of spin polarization (about\,30\% and 28\%) after optical irradiation, although the remagnetization is much faster in the case of Py/Au(100\,nm). For Py/MgO a quenching of about\,15\% was found. The corresponding OISTR-trace (OT), as the difference of the Ni and Fe signal, is shown in Fig.~\ref{fig:demags_and_OT_scaling}\,(b). While the OT for Py/Au(100\,nm) is close to zero, the OT for Py/MgO and Py/Au(10\,nm) is similar, but slightly higher in the case of Py/MgO. One conclusion out of this is, that although the absorbed energy in the spin system is almost 50\% lower in Py/MgO than in Py/Au(10\,nm), the OT is still higher in Py/MgO.

	\vspace{-0.2cm}
For better comparison we have scaled the quenching of each sample to the value of 30\%, using Py/Au(10\,nm) as a reference (Fig.~\ref{fig:demags_and_OT_scaling}\,(c)) to linearly extrapolate the OT of Py/MgO with $\approx$\,30\% of quenching. 
This was done by first selecting a time window between \SI{644}{\femto\second} and \SI{991}{\femto\second} for Py/MgO, \SI{340}{\femto\second} and \SI{810}{\femto\second} for Py/Au(100\,nm). The averaged spin polarization in this window was then calculated to be $ 0.828$ for Py/MgO and $0.728$ for Py/Au(100\,nm). This value was then subtracted from the whole time-resolved curve, resulting in different normalized levels before time zero $l_{T_0}$, $0.165$ for Py/MgO and $0.265$ for Py/Au(100\,nm). Now the whole time resolved curve was multiplied by $0.3/l_{T_0}$ and summed to 0.7. The result is shown in Fig.~\ref{fig:demags_and_OT_scaling}\,(c), where all time resolved curves show similar quenching of $\approx$\,30\%. We then extract the OT of each sample, as shown in Fig.~\ref{fig:demags_and_OT_scaling}\,(d). Now the OT of Py/MgO is about a factor of two larger than Py/Au(10\,nm), showing the trend of decreasing OT with increasing Au-thickness.

	\section{Theoretical model}
	
	We used a model based on full Boltzmann collision integrals to calculate the dynamics of the electronic distribution, which we used to determine the time-dependent occupation of states as described in the main text. The full details of the model are described in Ref.~\cite{Mueller2013PRB}.
	To determine the absorption profiles for the considered scenarios, we solved the Helmholtz equation for Al$_2$O$_3$(2\,nm)/Py(10\,nm)/Au(d)/MgO(500\,nm) heterostructures with $d = \SI{10}{\nano\meter}$ and $d = \SI{100}{\nano\meter}$, respectively. Details of the calculations can be found in the supplementary information of Ref.~\cite{Seibel2022}. 
	The refractive index for Al$_2$O$_3$ was taken from Ref.~\cite{Boidin2016}, for Py from Ref.~\cite{Tikuivsis2017}, for Au from Ref.~\cite{Werner2009} and for MgO from Ref.~\cite{Stephens1952}. For all values, a wavelength of \SI{800}{\nano\meter} was considered in accordance with our experiments. We determined the absorbed energy density by integrating the absorption profiles in the respective materials and considering the experimental fluence of \SI{28.1}{\milli\joule\per\centi\meter\squared}. The intensity of the laser excitation was then adjusted in the model to reproduce the energy absorbed in the experiment. 
	For the two limiting cases of no transport and full ballistic transport in the case of the \SI{100}{\nano\meter} Au layer, we took into account only the first \SI{10}{\nano\meter} of the absorption profile for the former case and the full \SI{100}{\nano\meter} for the latter case. 
	The resulting averaged fractions of absorbed energy are \SI{0.36}{\percent\per\nano\meter} for the \SI{10}{\nano\meter} Au layer, \SI{0.03}{\percent\per\nano\meter} for the \SI{100}{\nano\meter} Au layer and \SI{0.15}{\percent\per\nano\meter} for the limiting case of no transport in the \SI{100}{\nano\meter} Au layer.